\documentclass[12pt,a4paper]{article}
\usepackage{amssymb,amsmath,amstext,array,cite,graphicx,a4wide,ae}
\usepackage[latin1]{inputenc} 
\usepackage{t1enc} 
\usepackage{color}
\begin{document}
\thispagestyle{empty}

\mbox{}

\vspace{1cm}

\begin{center}

\LARGE\bf 
  Malleable parallelism with minimal effort\\ for maximal throughput and 
  maximal hardware load.\\[3cm]
   \Large\it F.~Spenke$^\ast$, K.~Balzer$^\dagger$, S.~Frick$\ddagger$, 
             B.~Hartke$^\ast$, and J.~M.~Dieterich$^+$  
             \\ [1cm]
   \large\rm $\ast$ Institute for Physical Chemistry,\\
             Christian-Albrechts-University,\\
             Olshausenstr.~40,\\
             24098 Kiel, GERMANY\\[3mm]
             $\dagger$ University Computing Center,\\
             Christian-Albrechts-University,\\
             Ludewig-Meyn-Str.~4,\\
             24118 Kiel, GERMANY\\[3mm]
             $\ddagger$ Zuse Institute Berlin (ZIB),\\
             Takustr.~7,\\
             14195 Berlin, GERMANY\\[3mm]
             $+$ Mechanical and Aerospace Engineering,\\
             Princeton University,\\
             Princeton, NJ 08544-5263, USA\\[5mm]
\end{center}

\newpage

\thispagestyle{empty}

\section*{Abstract}

In practice, standard scheduling of parallel computing jobs almost always
leaves significant portions of the available hardware unused, even with
many jobs still waiting in the queue. The simple reason is that the resource
requests of these waiting jobs are fixed and do not match the
available, unused resources. However, with alternative but existing and
well-established techniques it is possible to achieve a fully automated,
adaptive parallelism that does not need pre-set, fixed resources. Here, we
demonstrate that such an adaptively parallel program can 
run productively on a machine that is traditionally
  considered ``full'' and thus can indeed fill in all
such scheduling gaps, even in real-life situations on large supercomputers
in which a fixed-size job could not have started.

\vspace{2cm}

\subsection*{keywords}
  adaptive parallelism, malleable parallelism, scheduling, genetic algorithms,
  non-deterministic global optimization

\newpage

\setcounter{page}{1}

\section{Introduction}

Traditional job scheduling for high-performance computing (HPC) machines
requires fixed, pre-set resources (amount of memory, number of CPUcores or
compute nodes, maximum required time) for each job. These settings are provided
by the user, and most of them cannot be changed after job submission, neither by the scheduler
nor by the user. With a realistic job mix in real-life situations, this leads
to a total machine load being substantially smaller than 100\%, in at least two
generic situations:
\begin{enumerate}
\item during standard operation, $N$ compute nodes remain unused because the
  smallest jobs in the queue requests $M>N$ nodes; hence, typically, an HPC
  installation is considered ``full'' already at average loads of 90\%;
\item before huge jobs requesting a big portion (50\%-100\%) of the whole
  machine can run, a correspondingly big portion of the machine has to be
  drained completely; during this time, no new jobs can start that request
  more time than the time left until the scheduled start of the big job;
  typically, this can lead to the load dropping down towards zero, for
  extended times of several hours.
\end{enumerate}
Idle computing resources are not productive in terms of producing results, but
they still cost real money. Electricity consumption of the actual hardware and
of the periphery (cooling) may be smaller in idle mode, but is not zero -- and
should not be zero: In scenario (2) mentioned above, if the huge job finally
starts, heat dissipation is instantly required, and then the cooling should
already be up and running. 

As the size of installations grows, the amount of computing performance left
over under these conditions is non-negligible and may in itself be sufficient 
to run computationally intensive large-scale simulations. Hence, parallel jobs
that can be adapted in their resource consumption during runtime 
can not only use otherwise idle hardware but also allow jobs
to start as soon as small amounts of hardware become available for short
times, and to grow as soon as more hardware becomes available --- in
contrast to fixed-size jobs that need to wait until all requested resources
are available for the requested time slot. Hence such malleable jobs
are highly desirable to fit into this reality, from the
perspectives of both the HPC users and the HPC centers.

In a limited way, \textit{moldable} applications
\cite{Cirne,Srinivasan,Rauschmayr} already 
approach some of these goals. For a moldable job, resources are not given as
fixed numbers but as ranges or as set of alternatives;
actual numbers selected from these possibilities are set 
at the start of the otherwise traditional, fixed-size application. The only
prerequisite is that the application tolerates to be executed with more than
one amount of hardware resources. Clearly, under favorable circumstances,
this may allow a job to start earlier but still to complete all calculations
by running longer. 
Note that selecting one of several possible resource settings before
submitting a job to a queueing system is what most HPC users do
manually. In contrast, moldability is about shifting this choice to a slightly
later stage (just before execution, and while the job is already in the queue)
and/or about automating it, which requires minor or major adjustments in the
scheduling/queueing system. 
In any case, however, once such a job has started, it
cannot adapt to changing resource availability at any later point during its
runtime --- but such changes are happening continuously in any multi-user
environment. In contrast, this work addresses malleability or adaptivity,
i.e., exactly the feature of adapting to changing resource availability
during runtime.

In the past few years, there has been intensive research on the topic of
scheduling malleable parallel applications \cite{Jansen,Blazewicz,Fan,Cao},
but actually enabling malleability at the application level has lagged behind,
as already diagnosed by other authors \cite{Lemarinier}, and has been confined
to the computer-science (CS) literature so far. 

Some of these CS studies employed libraries that realize malleability by a
brute-force sequence of checkpointing, migrating and restarting the parallel
job in question \cite{Varela,Lemarinier}, implemented on top of MPI-2. 
Others employ more intricate methods, e.g., the ReShape framework
\cite{Sudarsan2010,Sudarsan2016}, also built on-top of MPI-2, or the flexMPI
framework \cite{Carretero2015,Carretero2016} built on MPI-3, or the proposed
ULFM standard API \cite{Lemarinier} that perhaps will be included in the
future MPI-4. Yet other studies \cite{malleable,Ribeiro,Iserte,Kale} employed
various application management systems (Charmm++, EasyGrid, Nanos++) as
additional software layer between the application and the scheduler. In all
cases, significant instrumentation of the application source code was needed,
to add defined "resize points", to describe data dependencies, to handle data
migration, and to achieve communication with the scheduler. Likewise, in most
cases the additional software between application and scheduler also had to be
modified/extended, to accomodate this scheduler-application communication and
to make resizing decisions. The applications employed in these studies ranged
from artificial "sleep"-jobs via basic algorithm parts also occuring in
real-life programs (e.g., Jacobi or conjugate-gradient algorithms) to small,
simplified versions of real-life applications (e.g., simple N-body
dynamics). Tests were typically performed on dedicated hardware, against a 
backdrop of artificial other jobs, with real-life-like characteristics.

One study \cite{Sudarsan2016} aimed at applying a ReShape-like approach to the
real-life molecular-dynamics program LAMMPS \cite{lammps}. In that case, however, the
authors felt the need to limit the necessary source-code changes. Hence, they
stepped back from the actual ReShape to employing the checkpoint/restart
feature already built into LAMMPS; but even then additional source-code
changes were required (including modifications of the input parser). 

In contrast to the CS papers listed above, we have employed a truly real-life
application, namely our global optimization code \textsc{ogolem}
\cite{ogolem}, which has been thoroughly benchmarked
\cite{bench,ogoreaxff,rmi,ljlarge} 
and has been used on very diverse, real-life global optimization problems
\cite{kana,ljmix,waterpersp,disulfide,golps}
However, other than in Ref.~\cite{Sudarsan2016}, we did not resort
to a brute-force checkpoint/restart approach to achieve malleability; this
feature arises as a side-effect of the
remote-method-invocation-(RMI)-parallelization of \textsc{ogolem} 
which we recently presented \cite{rmi}. 
Of course, we originally did have to do some source-code-level integration into \textsc{ogolem}
as subsequently discussed, but
only to introduce the RMI-parallelization itself. Notably, none of the computational kernels needed to be changed and the overall global optimization algorithm remained the same. As already described in
Ref.~\cite{rmi}, RMI is a simpler and more versatile parallelization basis
than MPI, allowing wildly heterogeneous hardware and full fault tolerance
\cite{rmi} without sophisticated add-ons.
No additional source-code modifications were necessary to get
from this RMI stage to the present malleability demonstration. In particular,
and again in contrast to the above CS studies, to achieve this malleability no
additional software layer was needed between \textsc{ogolem} and the scheduler; the
simple reason for this is that in our case no direct communication is needed
between the application and the scheduler, whereas in the above-cited studies
a direct, two-way communication is essential. In fact, both the hardware fault
tolerance demonstrated earlier \cite{rmi} and the communication-free
malleability demonstrated 
here arise from the same basic feature: In our RMI-based, asynchronous
server-client scheme, the server tolerates sudden loss and addition of
clients at any point during the overall calculation, without the need to
insert any resize points or data re-shuffling operations into the application
software. Hence, the resources allocated to this whole server-client setup can
be set externally by the scheduler, without the need to communicate any resize
events to the application level.
In fact, in our implementation, the external dependencies are minimal. 
We almost exclusively rely upon standard components of the Java virtual
machine; only the interaction with the scheduler has to be adapted to the
actual scheduler present. This clearly maximizes portability.
With several application examples on different HPC
installations, we show that such transfers between different
scheduler/queueing systems are easily possible.
Furthermore, we have run our tests and demonstrations not on dedicated 
HPC hardware sections and against a background-load of artificial jobs, but on
standard HPC installations, amidst an everyday job mix generated by all the
other users that happened to be present.

Last but not least, our overall aim is different: The CS studies summarized
above typically aim at optimizing aims like walltime or hardware energy
consumption, for a single malleable job in question, or for several
such jobs. Such aims require empirical testing and/or automated decisions
involving non-trivial performance forecasts based on collected performance
data, in short, they again require considerable additional algorithmic work in
an additional software layer. In contrast, we aim at minimizing overall
hardware idle time, and at using those otherwise idle resources productively.
Importantly, we can achieve this without any additional software,
relying solely on a few simple settings and on standard scheduler features.

In the applied scientific computing on today's HPC hardware, malleably or
adaptively parallel jobs simply do not occur. Instead, traditional
thread-based and plain MPI-1/2-based parallelization is the exclusive standard.
All jobs are run with fixed resource allocations, pre-set by the users,
which leads to the above-mentioned two standard situations, both of which
waste considerable HPC resources every day. Additionally, no out-of-the-box fault tolerance
is present, resulting in jobs on thousands of nodes being killed if only one
of them suffers a hardware failure. Our parallelization strategy described
here, and also those mentioned above, can change all that. To promote such
paradigm changes in the community of applied scientific computing, we
deliberately publish this Article not in a CS journal but in a theoretical
chemistry journal.   

The paper is organized as follows. Our new malleable parallelization strategy
is described in section~\ref{sec:alg}, the main experimental
results are in section~\ref{sec:experiments}, and the conclusions follow in
section~\ref{sec:conclusions}.

\section{Malleable parallelization}
\label{sec:alg}

For the original, detailed discussion of our RMI-based resilient parallelization,
we refer to Ref.~\cite{rmi}. We hence will restrict ourselves in the following
to a short discussion of the algorithmic features and details of relevance to
this work.

A server process maintains an internal queue of tasks to be accomplished, is
responsible to receive intermediate results, and combines them to the final
result. This process requires only very little resources, so it is often started
on infrastructure nodes.
Any number of client processes can attach at any point to the server
process to obtain a list of tasks to work on and, upon completing them,
return the intermediate results to the server. Server and clients maintain
heartbeat connections to assure that client malfunction from, e.g., hardware
failures are gracefully handled by the server and server shutdowns cause all 
clients to shutdown reliably on their own. Excellent parallel scalability to
6144 cores for evolutionary algorithms was demonstrated.\cite{rmi}

Obviously, the ability to attach and detach client processes trivially at job
runtime lends itself naturally to adapt a job to available computing resources,
i.e., be used as a \emph{fill-in} job.
We can hence use scheduling leftovers of ordinary, statically shaped jobs in an
efficient manner. As resources become available, clients are started and attach
to the server. As resources are needed for static jobs, clients can
and will be killed by the queuing system.

To achieve this goal of a complete machine load fill, the fill-in jobs need to be treated differently by the queuing system than ordinary jobs.
For an undisturbed workflow of ordinary jobs it is essential to cancel running fill-in jobs to free computing resources.
Luckily, this is a standard feature supported by modern scheduling systems.
Without excessive, costly, and usually inflexible checkpointing, an ordinary
job suffers substantial loss of computing time on a sudden cancellation, so this
feature often remains unused. In contrast, in our setup, the master-client
dialogue is built on task/result chunks that are communicated frequently and
can be controlled in their size and frequency by the user. Additionally, lost
computing time due to client cancellation can be rescheduled or, if possible
(as in the present evolutionary algorithm (EA) applications), just dropped. Hence, losses upon
cancellation can be kept minimal, and the balance between chunk size,
communication frequency and computations that possibly have to be re-done
can transparently be adapted
by the user to an expected interruption frequency scenario.

In contrast to the other approaches mentioned in the introduction, our
strategy does not require any direct cooperation between the scheduling system
on the one hand and the malleable server-client application described above.
In fact, the scheduler remains completely unaware of the presence of malleable
jobs; it only sees the presence of higher- and lower-priority jobs.
The scheduler is free to terminate these lower-priority jobs (RMI clients in
our case) at any point without notice,
as our approach will internally recover from said termination.
Thus, from the viewpoint of the traditional, statically
sized, high-priority jobs, there is no difference between free resources and
resources used by lower-priority fill-in jobs; the latter are always just as
easily and instantly available as the former.

Similarly, more clients can be added to the queue at any point, but again
without pre-planning and without app-scheduler communication. If
higher-priority jobs leave resources unused, the scheduler will start these
clients and they will then productively contribute to the overall
RMI server-client system. However, should no unused resources be available
currently, the clients simply keep waiting in the queue, without affecting the overall RMI
server-client job, except for a reduction in current throughput.
This total absence of direct scheduler-app communication not only simplifies
the software stack but also significantly reduces
error sources arising from mandatory cooperation between application / library
and scheduler. The only additional but very small, simple and dependency-free
tool needed is a way to re-supply the queue with new waiting client jobs, in a
reasonable proportion to the number of clients that transition from waiting to
active. This can either be a non-sophisticated shell script that keeps track
of the number of clients waiting in the queue, or a queue-script solution that
puts a new client into the queue once its own client job gets started.

In order to employ realistic fill-in jobs, we performed global minimum-energy
structure optimizations of water clusters with them, using the TIP4P
\cite{tip4p} water model, in most cases for a cluster size of (H$_2$O)$_{55}$,
which is the upper size limit for this system and for this task, in studies
published so far \cite{thakkar}.

\section{Experimental results}
\label{sec:experiments}

\subsection{Local hardware}
\label{subsec:local}

\begin{figure}[htbp]
  \centering
  \includegraphics[width=0.75\textwidth]{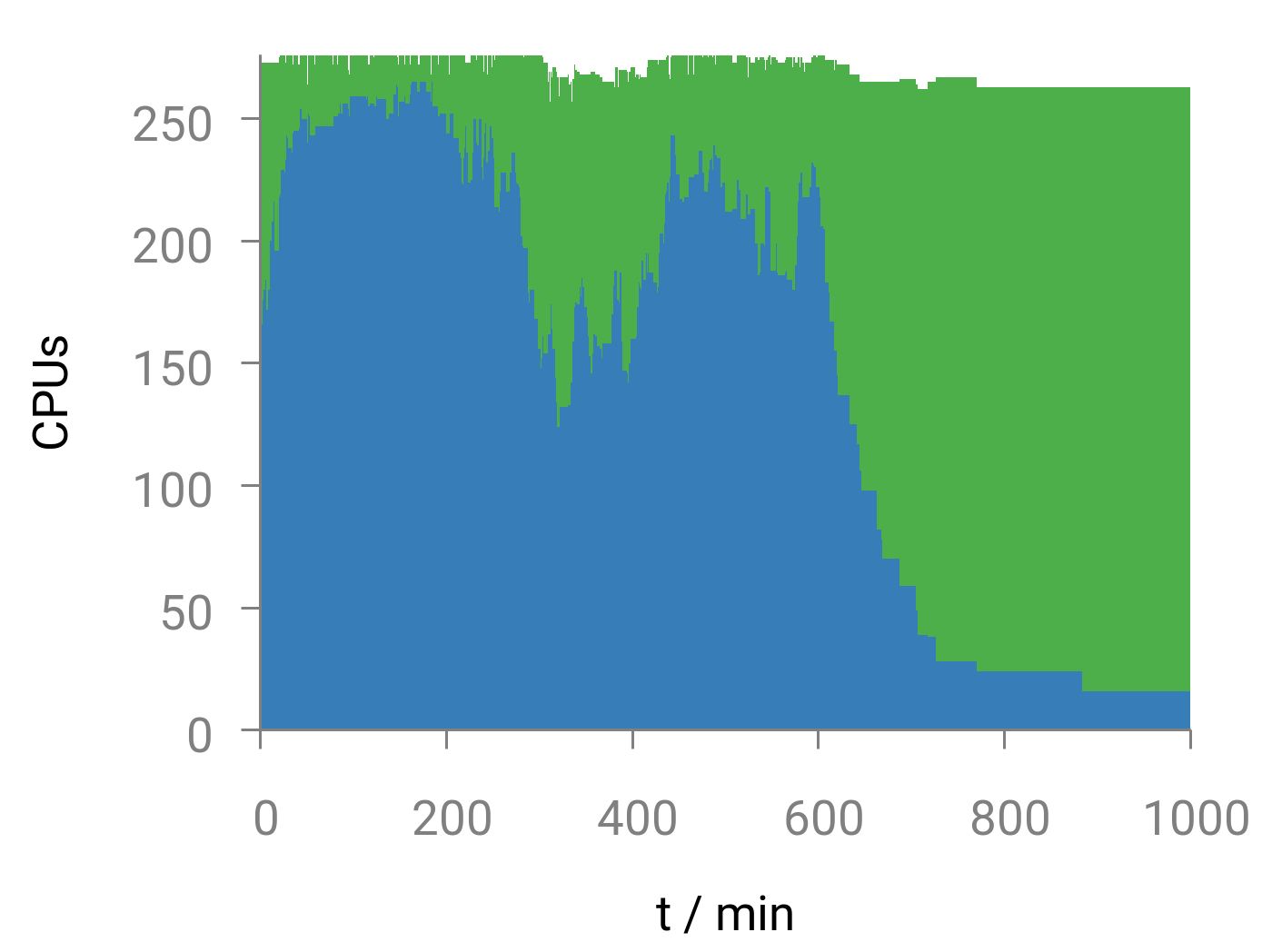}
  \caption{Queue fill-in at the Hartke workgroup computing cluster;
    normal jobs with fixed resources in blue, RMI-backfill-jobs in green.
    Note that the normal job set (blue) contains a random admixture of many
    short-lived jobs, leading to strong, irregular, high-frequency load
    oscillations. Nevertheless, the RMI-backfill is able to keep the overall
    load close to 100\% at all times.}
  \label{fig:AKload}
\end{figure}

As a first real-life test, we checked the queue-fill-in capabilities of our
RMI setup on the heterogeneous local hardware of the Hartke group. The
following machines were involved in the test shown in Fig.~\ref{fig:AKload}:

\begin{tabular}{cccc}
\# nodes & \# processors & processor type & \# cores / processor \\
4 & 2 & AMD Opteron 2358 SE & 4 \\     
2 & 2 & Intel Xeon E5-2680 v4 & 14 \\  
3 & 2 & Intel Xeon X5675 & 12 \\       
2 & 4 & AMD Opteron 6172 & 12 \\       
1 & 2 & Intel Xeon X5355 & 4 \\        
1 & 2 & AMD Opteron 2427 & 6 \\        
1 & 2 & Intel Xeon Gold 6154 & 18 \\   
\end{tabular}

and a login node with 8 processors of type Intel Xeon E5-2620.
All nodes run openSuse~42.x.
For all calculations a Java Runtime Environment (JRE) version 1.8.0\_73 was used.
The nodes are connected via 2x 1 GBit/s LACP.

The scheduling system used for batch processing is SLURM. \cite{slurm}
It is configured to distribute the existing resources on a first come, first serve basis.
To accommodate the fill-in jobs, a new partition was introduced.
Jobs started in this partition are low priority, i.e., they are only eligible to start
if no job in another partition can use the existing free resources.
Also these low-priority jobs are pre-emptible, i.e., they can be aborted
by the scheduler before they reach their currently set walltime.
The scheduler regards the resources occupied by jobs in this low-priority,
pre-emptible partition
as free for jobs in other partitions. If these resources are needed, 
low-priority jobs are canceled and subsequently requeued by the scheduler.

To test the adaptability of the backfill-jobs against a strongly
varying background of normal (fixed-resource) jobs, numerous 
short-lived jobs were added to the latter category.
The normal jobs running during the test consisted of photodynamics simulations with MOPAC
and global optimizations with \textsc{ogolem} (in shared memory mode).
The RMI-server was started on the login node. 
To variably fill both smaller load gaps but also the maximally available 276
cores with a compromise between high granularity and low job count, 40, 20 and
30 RMI-clients of size 1, 4 and 8 CPU cores were used as backfill-jobs.
None of the normal jobs was hindered or affected in any way by the backfill
jobs, and no manual guidance of the fill-in was made. 
Fig.~\ref{fig:AKload} shows a typical resulting breakdown of the total
CPU load into these two categories (normal jobs, fill-in jobs).
Recording of the CPU load started immediately after the first jobs were
submitted and continued for 1000 minutes.

Obviously, despite a strongly varying load of standard jobs, the fill-in
automatically tops off the available hardware to maximum load, at all times.

\subsection{University computing center}
\label{subsec:CAU}

A similar queue-fill-in was demonstrated
on a homogenous subcluster of the university computing center.
This cluster consisted of 
7 nodes with 40 processors of type Intel Xeon E7-4820 
with 10 cores each 
(i.e., 280 CPUcores overall)
and a login node with 8 processors of type Intel Xeon E5-2640.
All nodes run CentOS Linux 7.x.
For all calculations a JRE version 1.8.0\_101 was used.
The nodes are connected via InfiniBand.

The used scheduling system is SLURM.
Changes to the scheduler are nearly identical to those described in section~\ref{subsec:local}.
There are again low-priority jobs that are started in a separate partition and can be canceled by the scheduler.
In contrast to the above-mentioned scheduler setup, the canceled jobs are not requeued.
A fair-share policy is used at this cluster to regulate the start of jobs competing
for the same resources. This does not impact the fill-in jobs since these are
low-priority and hence do not compete for resources. 

\begin{figure}[htbp]
  \centering
  \includegraphics[width=0.75\textwidth]{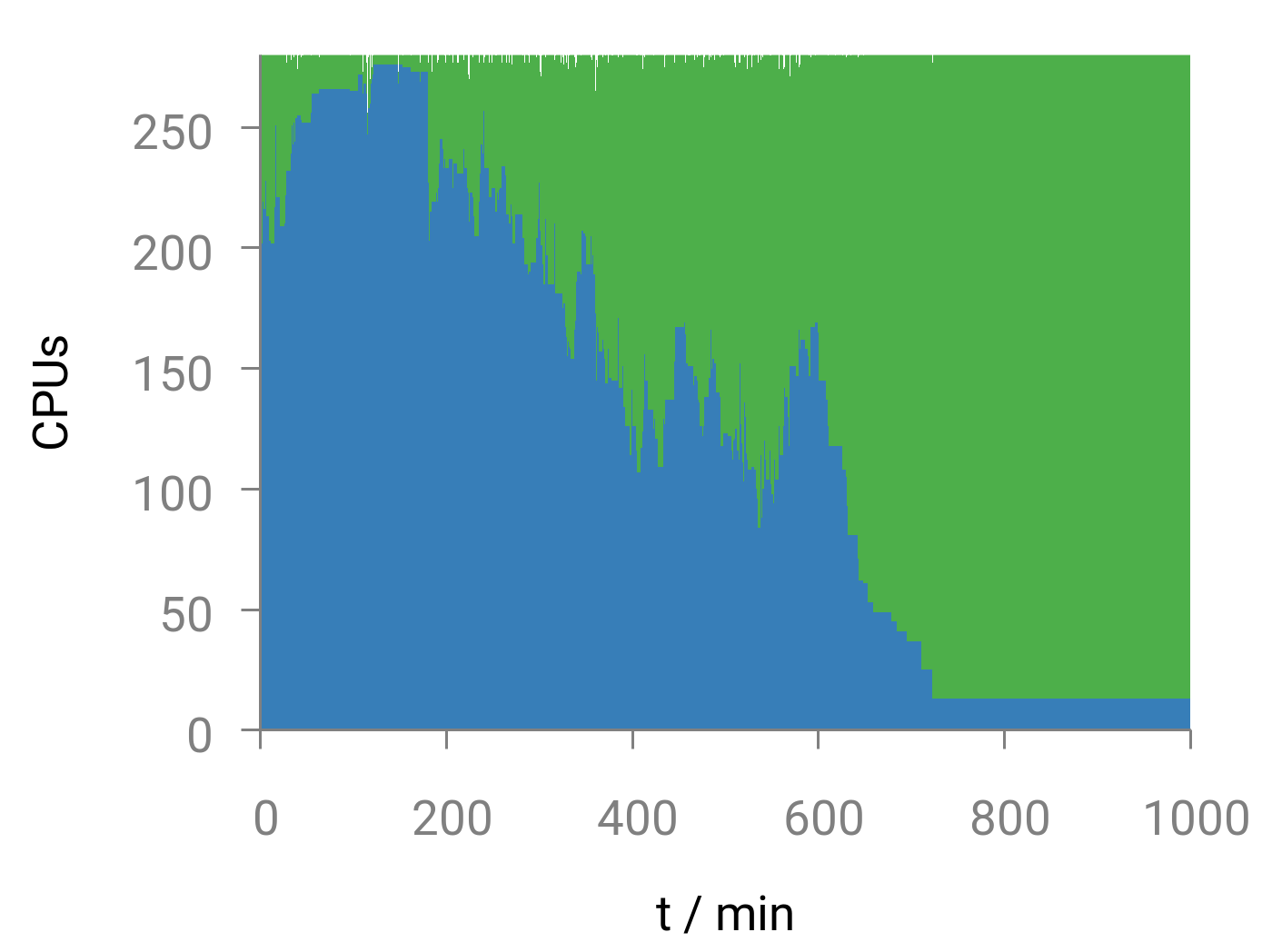}
  \caption{Queue fill-in at the CAUcluster;
    normal jobs with fixed resources in blue, RMI-backfill-jobs in green.}
  \label{fig:CAUload}
\end{figure}

The attempt to test the backfill-jobs at this hardware was again accompanied by
an artificial background of short-lived jobs in the main partition of the scheduler.
The RMI-server was started on the login node. 
Scripts were used to keep at least 5 backfill-jobs of size 1, 4 and 8
CPU-cores each queued and eligible to start, to be able to fill upcoming load
gaps of various sizes with a fairly small number of fill-in jobs.
Fig.~\ref{fig:CAUload} shows the CPU load during the test.
Again the CPU load was recorded for 1000 minutes after the first jobs were
submitted. 

\subsection{Regional HPC hardware}

A core application case of this work is a demonstration that
our flexible RMI parallelization
allows us to completely fill even the biggest load gaps on TOP500-class HPC
installations, during normal operation and without any impact on other
jobs. For such a demonstration, we have chosen the ``Konrad'' cluster in
Berlin, which is part of the North-German Supercomputing Alliance (HLRN)
\cite{HLRN} installations in Berlin and Hannover, serving hundreds of HPC
users in the North-German states Berlin, Brandenburg, Bremen, Hamburg,
Mecklenburg-Vorpommern, Niedersachsen and Schleswig-Holstein. In November
2017, it ranked 182 on the Top500 list.\cite{top500}

For the demonstration case in this subsection, we used the MPP1 section of
``Konrad'', a Cray XC30 consisting of 744 Intel Xeon IvyBridge compute
nodes. Each node contains 24 CPUcores (in two Intel Xeon IvyBridge E5-2695v2
2.4 GHz processors). Hence, this MPP1 section comprises a maximum of 17,856
CPUcores and has a theoretical peak performance of 342.8 TFlop/s.

As in the tests described above,
to allow our fill-in jobs at the HLRN, a new class for pre-emptible,
low-priority jobs was specified to the Moab scheduler running there.
According to HLRN policies, the number of jobs is restricted via class
settings, in our case to 24 simultaneously running jobs.

Additionally, we employed dynamically assigned resources:
During a fill-in run we submitted jobs in a per-minute interval that were tailored to fit
the available resources at that moment until the next reservation of a regular job.

Independent RMI-client processes were then started on each node.
The RMI-server process was started on one of the scheduling nodes of the HLRN.

The dynamically scaled fill-in jobs can easily fill any free nodes.
However, some restrictions arise due to HLRN user policies and scheduler limitations.
The optimal case would be one fill-in job per node.
In the event of a newly started ordinary job, exactly the needed number of
nodes could be freed and the new job could be started on these.
If the cancelled fill-in jobs were bigger than the new regular job, a portion of freed 
nodes will not be used and needs to be refilled with fill-in jobs.

To adapt to the limited job number according to HLRN user policies, we implemented 
an additional, automated job monitoring: 
If this limiting number was to be exceeded, e.g., because another regular job finished execution, 
the smallest running fill-in job was canceled and a new, larger one was
started on the combined free nodes. This once more exploits the high
adaptibility built into our concept.

\begin{figure}[htbp]
  \centering
  \includegraphics[width=0.75\textwidth]{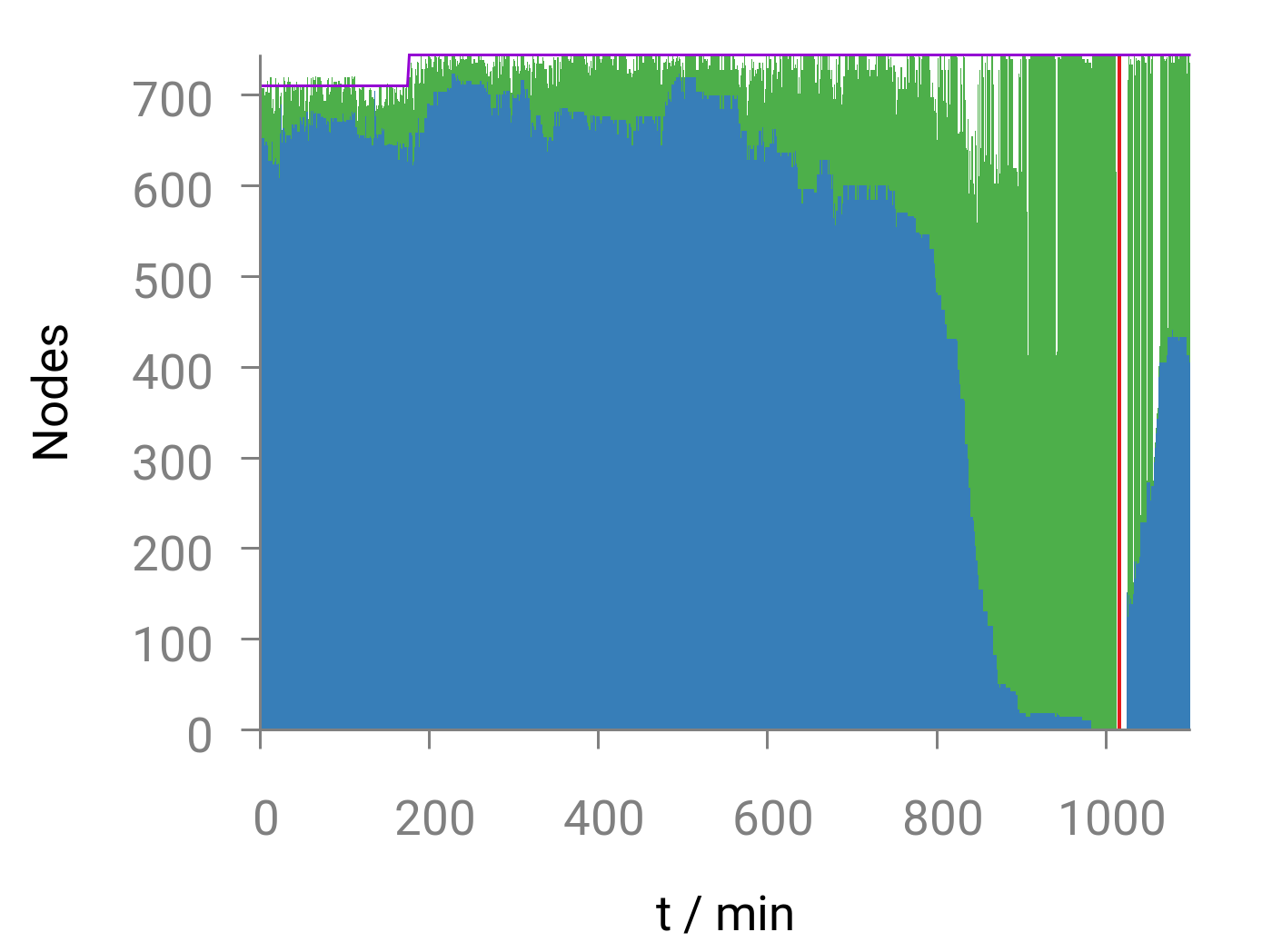}
  \caption{Queue fill-in at the HLRN computing center. A huge job (filling all
    of the machine) starts
  at t = 1010 minutes (visible only as narrow red line, because it finished
  again very quickly). In preparation for this
  event, normal job load (blue) decreases from about 100\% to 0\% between
  t = 500 minutes and t = 990 minutes, in a non-linear and unpredictable
  manner, since this decrease is not guided externally but arises
  spontaneously: The remaining time until the pre-set huge job start cannot be
  used by standard fixed-shape jobs available in the queue. However,
  our fill-in (green) covers this big gap completely, as it already did
  during normal operation (before t = 500 minutes), where the overall load
  also frequently failed to reach 100\%.
  The blue line indicates the maximally available resources at all times. They
  were not completely constant: Until t$\approx$200 minutes, a part of the
  system  was reserved for administrative purposes. Note that 1 node contains
  24 CPUcores, hence 744 nodes correspond to 17,856 CPUcores.} 
  \label{fig:HLRNload}
\end{figure}

Figure~\ref{fig:HLRNload} illustrates this fill-in on the HLRN hardware, both during
normal operation and during the machine draining period before a huge,
machine-wide job starts, i.e., this test covered both scenarios mentioned at
the beginning of the introduction. 
The CPU load was recorded for 1100 minutes in one-minute
intervalls, starting after the first batch of jobs was submitted.
Additionally,
from t = 0 to t $\approx$ 200 minutes, a portion of the nodes was reserved for
test jobs by the HLRN, 
so the maximally fillable node count is slightly lower than during the
remaining time. However, our setup is sufficiently flexible to also cope with
such changes, without manual interventions.

In contrast to the two test cases described in the previous subsections
\ref{subsec:local} and \ref{subsec:CAU}, at the HLRN we had no influence
whatsoever on the normal jobs (blue load in Fig.~\ref{fig:HLRNload}). Instead,
these jobs had been submitted by other regular HLRN users, as in any other
period of normal HLRN operation. Hence, this third test case at HLRN not only
demonstrates our adaptive fill-in at a very large HPC center but also under
perfectly normal, real-life conditions.

The reactions of our adaptive fill-in to current load levels and their changes
are essentially instantaneous, but reaction times of the HLRN queueing system
(including simple load-level queries, as needed to produce this load figure
and to drive the deployment of our fill-in jobs)
are not negligible, since high-frequency scheduling/queueing queries were not
part of the design specifications when this queueing system was set up several
years ago. These delays lead to the jagged appearance of the upper edge of the
fill-in load (green area in Fig.~\ref{fig:HLRNload}), occasionally escalating
into highly oscillatory load dips (between t=800 and t=900 minutes) or white
``stripes'' across the whole (green) fill-in load (between t=1000 and t=1100
minutes, after the big job finished). These irregularities were much smaller
in the other two demonstrations (Figs.~\ref{fig:AKload} and
\ref{fig:CAUload}). Hence, they are artifacts from a limited scheduler time
resolution at HLRN, and do not reflect true deficits of our own
fill-in setup. Conversely, to fully exploit these fill-in capabilities, the
design specifications of an HPC scheduler installation should include
sufficient responsiveness to higher-frequency queries than needed in more
traditional HPC scheduling.

With the fill-in jobs displayed in Fig.~\ref{fig:HLRNload}, over 75~M global
optimization steps for a (H$_2$O)$_{55}$ cluster in the TIP4P model could be
performed and 80~k CPU-hours were used. 

Most of these 80~k CPU-hours accrued between $t = 800$ min.~and $t = 1000$
min., i.e., within only 3 hours and 20 minutes. Nevertheless,
this is a substantial HPC resource usage, equivalent to what one full,
typical HLRN project uses within one whole week. With conventional parallel
(or serial) jobs, these 
resources would have remained idle, i.e., wasted, as evidenced by the mere
existence of all these load gaps and by the obvious
fact that normal HLRN jobs were not able to fill them. The HLRN
queues were never empty during this time, but typically contained 50--100 jobs
in a submitted/waiting state, 
and none of these normal, waiting jobs was in any
way hindered by our fill-in jobs,
since by construction the former have absolute priority over
the latter at all times.

\section{Conclusions}
\label{sec:conclusions}

In summary, we have shown that our highly flexible RMI-parallelization of
\textsc{ogolem} \cite{rmi} can indeed be employed to adaptively fill in each and every
bit of computing resources that standard parallel jobs under standard
scheduling have to leave unused and to use them productively --- no matter if these leftovers are small or
huge. We have demonstrated this with one and the same program package on three
very different computer systems, with significant differences in the
queueing/scheduling software and with huge differences in hardware size,
ranging from a small, heterogeneous, local computing cluster to a national
Top500 supercomputer. To transfer our setup from one of these machines to
another one, no changes at all were necessary in our \textsc{ogolem} package, and only
minor adaptations had to be made in small helper scripts (interacting with the
scheduler) and in the scheduler setup (in all cases exclusively exploiting
already existing scheduler features). Additionally, on the largest system
(HLRN) our demonstration also was a fully real-life case, against a backdrop
of standard jobs from many other users, completely beyond our control.

Therefore, with present-day adaptively parallel technologies, even on
large-scale HPC installations and in everyday situations, it is now
demonstrably possible to avoid all scheduling losses and to achieve a total
load level of 100\%, at all times, maximally exploiting the available
computing resources. Or, in other words, this approach can be
used to run HPC jobs with substantial resource needs on HPC machines that are
``full'' from the perspective of traditional, fixed-size jobs.
While achieving 100\% machine load certainly is a desirable goal for any HPC
installation, it is not the only one possible with our scheme. With minor
additional modifications in the queue configuration or in the mechanism that
supplies further client jobs, the maximum allowed load for such an RMI-based
server-client job can also be set to arbitrarily smaller load
fractions. Again, this can be achieved without any changes in the
application software or in the RMI-server-client setup.

Of course, the RMI-parallel model described and demonstrated
here is not limited to our \textsc{ogolem} code or to GA-like algorithms. The basic
ideas that enable both the fail-safe heterogeneity shown in Ref.~\cite{rmi}
and the malleability shown here can be transferred to any algorithm that can
be set up in a fine-grained server-client model with limited data
exchange. Prime examples are all kinds of Monte Carlo (MC) calculations, in
chemistry including, e.g., MC-evaluations of thermodynamic quantities,
grand-canonical simulations of complex systems, coarse-grained kinetic MC
calculations (KMC) of reaction networks, or quantum-MC (QMC) evaluations of
the Schr\"{o}dinger equation, for nuclei or electrons. 
Another important application area is 
quantum chromodynamics, where MC-integrations are also heavily used. Such
calculations typically take 20\% of all CPU hours used at HLRN.
More generally, our
paradigm can be easily adopted by any algorithm that generates final results
by collecting data from many loosely coupled subtasks, i.e., also by methods
like parallel tempering or umbrella sampling molecular dynamics (MD). 
To enable transfer of the present strategies to even broader classes of
algorithms, it may be useful to combine not only thread-based and RMI-based
parallelization (as done here and in Ref.~\cite{rmi}) but to also include
MPI-like parallelization. Future work will include such
methodical advances as well as transfers to other types of calculations, as
discussed above.

\section*{Acknowledgments}
We would like to acknowledge Holger Marten of the Kiel University Computing
Center and Thomas Steinke, Christian Schimmel and the ``Fachberater''
team of the Zuse Institute Berlin (ZIB) / North-German Supercomputing Alliance
(HLRN) for allowing us to perform the real-life tests reported here on their
machines during normal operation, despite their non-standard
queueing/scheduling. Additionally, FS and BH are grateful for a computer time
grant which made the big fill-up calculations at HLRN possible, and to Peter
Hauschildt (Hamburg Observatory) for submitting huge astrophysics jobs at
HLRN, triggering machine drainings that we could then fill up.\\
JMD wishes to extend his gratitude to Scientific Computing \& Modelling (SCM)
who allowed him to pursue these questions in his free time. He also wishes
to thank Dean Emily Carter for her current and ongoing support of his other
scientific endeavours.\\ BH is grateful for funding by the Deutsche
Forschungsgemeinschaft DFG via project Ha2498/16-1.

\bibliographystyle{unsrt}

\begin{thebibliography}{10}

\bibitem{Cirne}
W.~Cirne and F.~Berman.
\newblock Using moldability to improve the performance of supercomputer jobs.
\newblock {\em J. Parall. Distrib. Comput.}, 62:1571--1601, 2002.

\bibitem{Srinivasan}
S.~Srinivasan, S.~Krishnamoorthy, and P.~Sadayappan.
\newblock A robust scheduling stratey for moldable scheduling of parallel jobs.
\newblock In {\em IEEE International Conference on Cluster Computing ({CLUSTER}
  2003)}, 2003.

\bibitem{Rauschmayr}
N.~Rauschmayr and A.~Streit.
\newblock Evaluating moldability of {LHCb} jobs for multicore job submission.
\newblock In {\em 15th international symposium on symbolic and numeric
  algorithms for scientific computing ({SYNASC 2013})}, 2013.

\bibitem{Jansen}
K.~Jansen.
\newblock Scheduling malleable parallel tasks: An asymptotic fully polynomial
  time approximation scheme.
\newblock {\em Algorithmica}, 39:59--81, 2004.

\bibitem{Blazewicz}
J.~Blazewicz, M.~Machowiak, J.~Weglarz, M.~Y. Kovalyov, and D.~Trystram.
\newblock Scheduling malleable tasks on parallel processors to minimize the
  makespan.
\newblock {\em Annals Oper. Res.}, 129:65--80, 2004.

\bibitem{Fan}
L.~Y. Fan, F.~Zhang, G.~M. Wang, and Z.~Y. Liu.
\newblock An effective approximation algorithm for the malleable parallel task
  scheduling problem.
\newblock {\em J. Paral. Distrib. Comput.}, 72:693--704, 2012.

\bibitem{Cao}
Y.~J. Cao, H.~Y. Sun, D.~P. Qian, and W.~G. Wu.
\newblock Scalable hierarchical scheduling for malleable parallel jobs on
  multiprocessor-based systems.
\newblock {\em Comput. Syst. Sci. Eng.}, 29:169--181, 2014.

\bibitem{Lemarinier}
P.~Lemarinier, K.~Hasanov, S.~Venugopal, and K.~Katrinis.
\newblock Architecting malleable {MPI} applications for priority-driven
  adaptive scheduling.
\newblock In {\em Proceedings of the 23rd European {MPI} Users' Group Meeting},
  EuroMPI 2016, pages 74--81, New York, NY, USA, 2016. ACM.

\bibitem{Varela}
K.~El Maghraoui, T.~J. Desell, B.~K. Szymanski, and C.~A. Varela.
\newblock Malleable iterative {MPI} applications.
\newblock {\em Concurrency Computat.: Pract. Exper.}, 21:393--413, 2009.

\bibitem{Sudarsan2010}
R.~Sudarsan and C.~J. Ribbens.
\newblock Design and performance of a scheduling framework for resizable
  parallel applications.
\newblock {\em Parallel Comput.}, 36:48--64, 2010.

\bibitem{Sudarsan2016}
R.~Sudarsan and C.~J. Ribbens.
\newblock Combining performance and priority for scheduling resizable parallel
  applications.
\newblock {\em J. Parallel Distrib. Comput.}, 87:55--66, 2016.

\bibitem{Carretero2015}
G.~Mart\'{\i}n, D.~E. Singh, M.-C. Marinescu, and J.~Carretero.
\newblock Enhancing the performance of malleable {MPI} applications by using
  performance-aware dynamic reconfiguration.
\newblock {\em Parallel Comput.}, 46:60--77, 2015.

\bibitem{Carretero2016}
M.~Rodr\'{\i}guez-Gonzalo, D.~E. Singh, J.~G. Blas, and J.~Carretero.
\newblock Improving the energy efficiency of {MPI} applications by means of
  malleability.
\newblock In {\em 24th Euromicro International Conference on Parallel,
  Distributed and Network-Based Processing}, 2016.

\bibitem{malleable}
A.~Gupta, B.~Acun, O.~Sarood, and L.~V. Kal\'{e}.
\newblock Towards realizing the potential of malleable jobs.
\newblock In {\em 21st International Conference on High-Performance Computing
  (HiPC)}, 2014.

\bibitem{Ribeiro}
F.~S. Ribeiro, A.~P. Nascimento, C.~Boeres, V.~E.~F. Rebello, and A.~C. Sena.
\newblock Autonomic malleability in iterative {MPI} applications.
\newblock In {\em 25th International Symposium on Computer Architecture and
  High-Performance Computing}, 2013.

\bibitem{Iserte}
S.~Iserte, R.~Mayo, E.~S. Quintana-Ort\'{\i}, V.~Beltran, and A.~J.~Pe\ {n}a.
\newblock Efficient scalable computing through flexible applications and
  adaptive workloads.
\newblock In {\em 46th International Conference on Parallel Processing
  Workshops}, 2017.

\bibitem{Kale}
S.~Prabhakaran, M.~Neumann, S.~Rinke, F.~Wolf, A.~Gupta, and L.~V. Kal\'{e}.
\newblock A batch system with efficient adaptive scheduling for malleable and
  evolving applications.
\newblock In {\em 29th IEEE International Parallel and Distributed Processing
  Symposium}, 2015.

\bibitem{lammps}
S.~Plimpton.
\newblock Fast parallel algorithms for short-range molecular dynamics.
\newblock {\em J. Comput. Phys.}, 117:1--19, 1995.

\bibitem{ogolem}
J.~M. Dieterich and B.~Hartke.
\newblock Ogolem: Global cluster structure optimization for arbitrary mixtures
  of flexible molecules -- a multiscaling, object-oriented approach.
\newblock {\em Mol. Phys.}, 108:279--291, 2010.

\bibitem{bench}
J.~M. Dieterich and B.~Hartke.
\newblock Empirical review of standard benchmark functions using evolutionary
  global optimization.
\newblock {\em Appl. Math.}, 3:1552--1564, 2012.

\bibitem{ogoreaxff}
M.~Dittner, J.~M\"{u}ller, H.~M. Aktulga, and B.~Hartke.
\newblock Efficient global optimization of {ReaxFF} parameters.
\newblock {\em J. Comput. Chem.}, 36:1550--1561, 2015.

\bibitem{rmi}
J.~M. Dieterich and B.~Hartke.
\newblock An error-safe, portable, and efficient evolutionary algorithms
  implementation with high scalability.
\newblock {\em J. Chem. Theory Comput.}, 12:5226, 2016.

\bibitem{ljlarge}
M.~Dittner and B.~Hartke.
\newblock Conquering the hard cases of {L}ennard-{J}ones clusters with simple
  recipes.
\newblock {\em Comp. Theor. Chem.}, 1107:7--13, 2017.

\bibitem{kana}
J.~M. Dieterich, U.~Gerstel, J.-M. Schr\"{o}der, and B.~Hartke.
\newblock Aggregation of {K}anamycin {A}: dimer formation with physiological
  cations.
\newblock {\em J. Mol. Mod.}, 17:3195, 2011.

\bibitem{ljmix}
J.~M. Dieterich and B.~Hartke.
\newblock Composition-induced structural transitions in mixed {L}ennard-{J}ones
  clusters: global reparametrization and optimization.
\newblock {\em J. Comput. Chem.}, 32:1377--1385, 2011.

\bibitem{waterpersp}
U.~Buck, C.~C. Pradzynski, T.~Zeuch, J.~M. Dieterich, and B.Hartke.
\newblock A size-resolved perspective of large water clusters.
\newblock {\em Phys. Chem. Chem. Phys.}, 16:6859--6871, 2014.

\bibitem{disulfide}
J.~M\"{u}ller and B.~Hartke.
\newblock A {ReaxFF} reactive force field for disulfide mechanochemistry,
  fitted to multireference ab-initio data.
\newblock {\em J. Chem. Theory Comput.}, 12:3913--3925, 2016.

\bibitem{golps}
B.~G. del Rio, J.~M. Dieterich, and E.~A. Carter.
\newblock Globally optimized local pseudopotentials for orbital-free density
  functional theory simulations of liquids and solids.
\newblock {\em J. Chem. Theory Comput.}, 13:3684--3695, 2017.

\bibitem{tip4p}
W.~L. Jorgensen, J.~Chandresekhar, J.~D. Madura, R.~W. Impey, and M.~L. Klein.
\newblock Comparison of simple potential functions for simulating liquid water.
\newblock {\em J. Chem. Phys.}, 79:926, 1983.

\bibitem{thakkar}
S.~Kazachenko and A.~J. Thakkar.
\newblock Water nanodroplets: Predictions of five model potentials.
\newblock {\em J. Chem. Phys.}, 138:194302, 2013.

\bibitem{slurm}
A.~B. Yoo, M.~A. Jette, and M.~Grondona.
\newblock {SLURM: Simple Linux Utility for Resource Management.}
\newblock In D.~Feitelson, L.~Rudolph, and U.~Schwiegelshohn, editors, {\em
  {Job Scheduling Strategies for Parallel Processing, JSSPP 2003}}, volume 2862
  of {\em Lecture Notes in Computer Science}. Springer, Berlin, Heidelberg,
  2003.

\bibitem{HLRN}
{North-German Supercomputing Alliance (HLRN)}.
\newblock \texttt{https://www.hlrn.de/}, accessed: 2017/11/24.

\bibitem{top500}
Top500 list.
\newblock \texttt{https://top500.org/list/2017/11/?page=2}, accessed:
  2017/11/24.

\end{thebibliography}

\end{document}